\documentclass{optica-article}
\journal{opticajournal} 
\usepackage{multirow,ulem}
\usepackage{array}
\begin{document}

\title{Complex refractive index measurements of Poly(methyl methacrylate) (PMMA) over the UV-VIS-NIR region}

\author{Pham Thi Hong,\authormark{1} Hung Q. Nguyen,\authormark{1} and H. T. M. Nghiem\authormark{2,*}}

\address{\authormark{1}Nano and Energy Center, VNU University of Science, Vietnam National University, 120401 Hanoi, Vietnam\\
\authormark{2}Phenikaa Institute for Advanced Study, Phenikaa University, 12116 Hanoi, Vietnam}
\email{\authormark{*}hoa.nghiemthiminh@phenikaa-uni.edu.vn}

\begin{abstract*}
Poly(methyl methacrylate), PMMA, is a popular polymer for optical applications due to its superior transmission and reflection. However, its optical properties in the ultraviolet regime still need to be thoroughly studied. Using the reflection-transmission method, we determine its complex refractive index by numerically analyzing the measured data from thin films. The PMMA standalone film is fabricated by peeling off its substrate after spin-coating. Its transmittance and reflectance are then measured in a spectrophotometer using an integrated sphere. The complex refractive index $n+ik$ is extracted theoretically from the measured transmittance $T$ and reflectance $R$. The uncertainties of measured $n$ and $k$ are discussed in the two limits of strong absorption and weak absorption of the materials to illustrate the advantages and disadvantages of the approach.
\end{abstract*}
% \keywords{Poly(Methyl Methacrylate), optical properties,  ultraviolet}

% \maketitle

\section{Introduction}
Poly(methyl methacrylate) (PMMA), also known as acrylic or plexiglass, has excellent mechanical strength and weather resistance \cite{Ali15}. It has superior optical properties in an extended range of wavelengths, spanning the ultraviolet (UV), visible, and infrared spectra \cite{nassier2022, Tuhin20, Beadie15, Zhang20}. Due to its high transparency, low scattering, and good resistance to UV radiation, PMMA is used in a broad range of applications, such as embedding inorganic particles in optoelectronic devices \cite{Tucu15,Hazim20}, electron transport layers in OLEDs \cite{Kandulna18,Nayak19}, selective filters in gas sensors \cite{Misha14, Hong15}, protecting layers in photonics devices \cite{Xu17 ,Bera19}, or radiative layers in radiative cooling applications \cite{Boulet14, Liu22,Qi22}. 
Despite the important role across widespread applications, the complex refractive index $n + ik$ of PMMA polymer is not sufficiently studied, especially in the ultraviolet region. To our knowledge, the real part $n$ is available in the ultraviolet \cite{Tuhin20} and visible to infrared region \cite{Kasarova07, Sultanova09, Languy11, Beadie15, Zhang20}. However, the imaginary part $k$ is only available in the visible and infrared range \cite{Tuhin20, Zhang20} and needs to be better studied in the ultraviolet region. 

The complex refractive index cannot be measured directly. Instead, it should be derived from other quantities, such as reflectances, transmittances, or angles of refraction for polarized light, by using a suitable theory. There are several methods to determine the complex refractive index of a polymer, for example, the Kramers-Kronig transform method \cite{zellouf1996,Srabo10,Silfsten11}, the spectroscopic ellipsometry \cite{Tuhin20}, or the reflection-transmission method \cite{Nichelatti2002,Brissinger19,Li17}. In the Kramers-Kronig transform method, the transmittances of thin films at different thicknesses are measured to extract the imaginary part $k$. The real part $n$ is derived from $k$ with the help of the Kramers-Kronig transformation. To improve accuracy, the transmittance should be measured with a broad spectrum \cite{Srabo10,Silfsten11}. 
 {In the spectroscopic ellipsometry method, the complex refractive indices are calculated from the change in polarized light upon light reflection on a sample \cite{Fujiwara07}. This method involves minimal human interaction, leading to reduced measurement errors and high sensitivity to very small changes in optical properties. Analysis models are selected to best fit to the materials of interest \cite{Tuhin20,Zhang20}. Similarly, in the reflection-transmission method, the complex refractive indices are calculated from measured reflectance and transmittance of film samples.} It has explicit analytic expressions to extract complex refractive index $n+ik$ of material from the measured transmittance and reflectance in the visible and infrared regions\cite{Nichelatti2002}. 
One should pay attention to the precision of the measurements when the sample transmittance is below the measurement sensitivity \cite{Brissinger19} or when it has strong transmission \cite{Li17}.
{It is worth noting that light interference in the infrared region may occur when the sample thickness is on the order of the wavelength \cite{Zhang1998,Sun2018}. It is necessary to employ appropriate models for this interference effect; however, this is out of our interest region. }

In this paper, we study the complex refractive index of Poly(Methyl Methacrylate) focusing on the UV regime to extract the optical constant $n+ik$ in the corresponding region. First, a thin PMMA film is fabricated, and its transmittance and reflectance are measured in a standard UV-Vis-NIR spectrophotometer in a full range from $0.2$ to $2.5$ $\mu m$ with the aid of integrating sphere. 
The reflection-transmission method is then utilized to solve for the complex refractive index where the light interference is excluded. We pay special care to the accuracy of our approach and compare our results to other works.

\section{Methods}
\subsection{Theoretical framework}\label{theory}
As an electromagnetic wave reaches and transmits through a finite-thickness film of a material, the reflectance $R$ and transmittance $T$ represent the power fractions of the incident light reflected by and transmitted through the film. The fractions are the accumulation of the infinite number of reflections between and transmissions through the two faces of the film. The theory is developed 
without light interference, which is applicable for the case of film thickness larger than the wavelength of the incidence light or the use of integrating sphere with highly-reflective diffuse coating on the inner surface \cite{Reule76, Kari92}.
 By using the ray-tracing procedure to track all the possible reflected and transmitted light rays \cite{Howell20}, $R$ and $T$ can be written as 
 {
\begin{align}
    R &=r+\frac{r(1-r)^2\tau^2}{1-r^2\tau^2},\label{eq:R0}\\
    T &=\frac{(1-r)^2\tau}{1-r^2\tau^2}, \label{eq:T0}
\end{align}
}
in which the reflection coefficient $r$ is a fraction of the light reflected at the air-film interface 
\begin{align}
      r &=\frac{(n-1)^2+k^2}{(n+1)^2+k^2}, \label{eq:r00}
\end{align}
when the light is at the normal incidence\cite{Bohren2008}. The {transmissivity $\tau$ } presents the power-loss factor as light propagates from one film surface to another
{
\begin{equation} 
    \tau = \exp{(-\frac{4\pi k}{\lambda}d)},\label{eq:t00}
\end{equation}
}
where $\lambda$ is the wavelength and $d$ is the film thickness.

Calculating the reflectance $R$ and transmittance $T$ from a given refractive index, $n+ik$, is straightforward by following equations~(\ref{eq:R0}-\ref{eq:t00}). On the contrary, the inverse problem of estimating $n$ and $k$ from the measured $R$ and $T$ requires more effort.

In this report, we calculate the complex refractive index from the measured data by using analytic expressions. From the relations between $T$, $R$ and {$\tau$}, $r$ in Eqs.~\eqref{eq:R0} and ~\eqref{eq:T0}, we suppress $r$ so that {$\tau$} is expressed by a quadratic equation 
{
\begin{equation}
    T\tau^2+[(R-1)^2-T^2]\tau-T=0.
    \label{eq:quad}
\end{equation} 
}
The solution to this equation is determined from the measured transmittance $T$ and reflectance $R$. It is
{
\begin{equation} \label{eq:t0}
      \tau=\frac{-[(R-1)^2-T^2]\pm\sqrt{\Delta}}{2T},
\end{equation}
}
with $\Delta=[(R-1)^2-T^2]^2+4T^2$. Here, the plus sign is chosen to guarantee the positive {transmissivity}. Using Eq.\eqref{eq:t00}, the extinction coefficient $k$ is given by
{
\begin{equation} \label{eq:k0}
      k=-\frac{\lambda \log(\tau)}{4\pi d}.
\end{equation}
}
From Eqs.~\eqref{eq:R0} and ~\eqref{eq:T0}, the reflection coefficient $r$ is
{
\begin{align}
      r&=\frac{2R}{1+\tau^2\pm\sqrt{(1+\tau^2)^2-4\tau^2R(2-R)}}, \label{eq:r0}
\end{align}
}
in which only the plus sign gives the relevant value of $r<R$. In consequence, $n$ is determined analytically from
\begin{align}
      n&=\frac{1+r}{1-r}\pm\sqrt{\frac{4r}{(1-r)^2}-k^2}. \label{eq:n0}
\end{align}
which is derived from Eq.~(\ref{eq:r00}) and the choice of plus sign is to ensure that $n$ is greater than $1$. 

Up to now, $n$ and $k$ are expressed analytically in relation to the other quantities. In reality, we numerically determine $n$ and $k$ from the measured $R$, $T$, and $d$. The final results contain possible errors inherited from the measurements and the measured sample. The instrument noise constrains the calculation of $k$ in a strong-absorption regime. 
When absorption is strong, Eqs. ~\eqref{eq:R0} and \eqref{eq:T0} can be approximated by
{
\begin{align}
    R &=r, \label{eq:R0A} \\
    T &=(1-r)^2\tau, \label{eq:T0A} 
\end{align} }
then Eq.~\eqref{eq:k0} is reduced into
\begin{equation} \label{eq:k0A}
      k=\frac{\lambda \log((1-R)^2/T)}{4\pi d} <\frac{\lambda \log(T^{-1})}{4\pi d}.
\end{equation}
{The lowest reliable transmittance is in the scale of instrument noise, which normally equals to $0.001 \%$ \cite{Brissinger19}.}
Hence
\begin{equation} \label{eq:k0A_2}
      k <\frac{5\lambda \log(10)}{4\pi d}.
\end{equation}
{That means, for a given sample-thickness, $k$ is always less than a finite value.} Therefore, the reflection-transmission method may fail to provide the correct extinction coefficient $k$ result when the film sample is too thick \cite{Nichelatti2002, Brissinger19}. To solve this, one should reduce the thickness of the sample to increase the upper bound on the right-hand side of Eq.~(\ref{eq:k0A_2}) so that the transmittance through the film sample is within the reliable range of measurement.

The reflection-transmission method presented here is similar to those in previous studies \cite{Nichelatti2002, Brissinger19} except for the order of estimating the reflection coefficient {and transmissivity} for the convenience of our calculations. 

\subsection{Sample preparation} \label{farbication}
\begin{figure}[htb]
    \centering\includegraphics[width=0.8\textwidth]{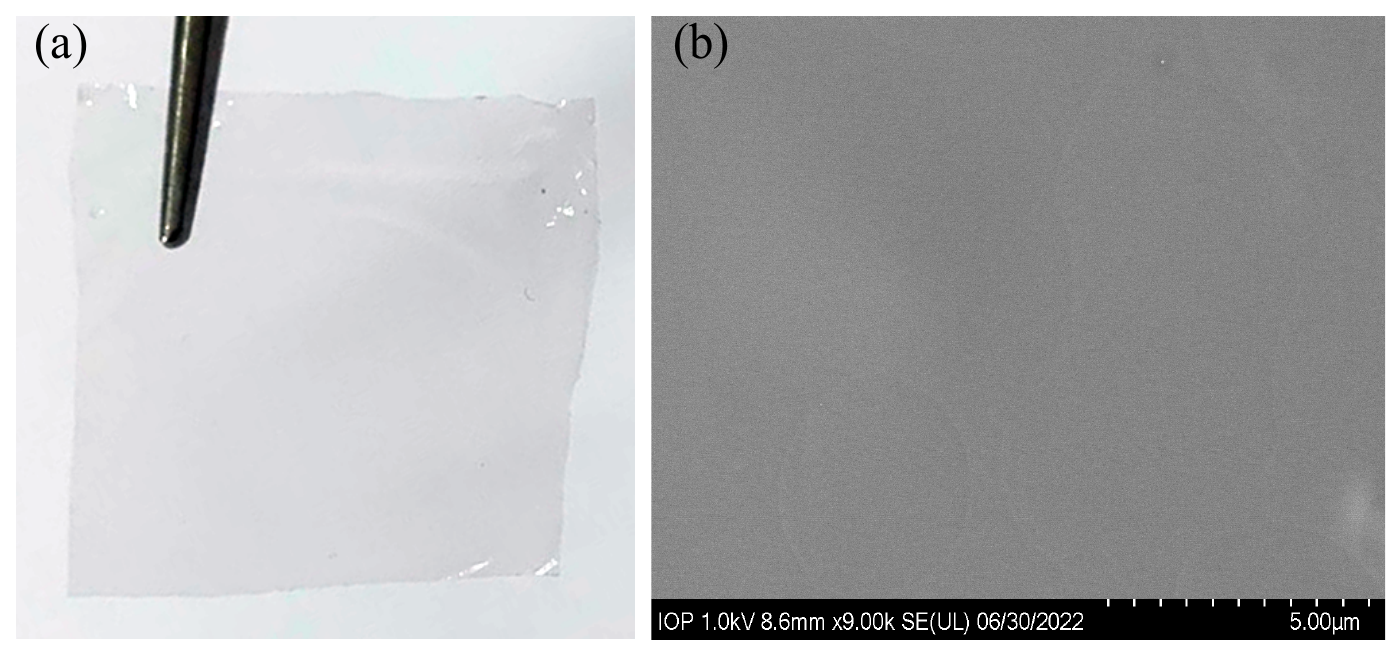}
    \caption{\textbf{Morphology of PMMA film:} (a) The picture of the sample after peeling off from the glass substrate with size $2\times2$ cm$^{2}$. (b) The SEM image of the film at a small scale. }\label{fig:morphology}
\end{figure}

Poly(Methyl Methacrylate), or PMMA, films are fabricated in a standard cleanroom using spin-coating. The PMMA obtained from Sumitomo Chemical Co. is in granular form with a molecular weight of $1.18$ g/cm$^3$ ~\cite{Gnanavel19}. It is dissolved in pure acetone with a ratio of $1:8$ at temperature $21.5^\circ$C  using magnetic stirring for $2$ hours. The solution is spin-coated at $300$ rpm in $3$ minutes using a standard spin-coater on a glass substrate with $2.54\times 2.54$ cm$^2$ in size. Since acetone evaporates quite fast, the sample dries on itself without any baking step. The sample is then peeled off from the glass substrate with the help of copper tape and then transferred into the spectrophotometric measurements. The standalone PMMA film image is shown in figure~\ref{fig:morphology}.a. It is mandatory to peel off the film from the glass substrate since the thick glass substrate dominates over the absorption property of the thin PMMA sample in the UV regime \cite{Olenych17}. 

The thickness of PMMA film is determined using direct and indirect measurement methods. In the direct measurement, the PMMA film on the glass substrate is measured using a KLA Tencor D-100 profilometer at five different locations on the film, shown in Table~\ref{tab:thickness}. The average thickness is $d \pm \Delta d = 5.8\pm0.9$ $\mu$m. In the indirect measurement, we determine the sample thickness by fitting the transmittance spectra at the local maxima of absorption ($\lambda\approx5.7$, $6.9$, $8.0$, $8.7$, and $10.1$ $\mu m$) obtained from the Fourier-transform infrared spectroscope to the published data \cite{Zhang2020b}. 
The FTIR measurements are done at five different positions, and the fitting results of the standalone PMMA thickness are shown in Table~\ref{tab:thickness} with the averaged thickness $d \pm \Delta d=6\pm0.6$ $\mu$m. The sample-thickness results obtained from the two methods are approximately equal. We use the sample thickness measured from the profilometer in the calculation of the optical properties of PMMA in the next section. Also, the homogeneity of the film is estimated on the scale of 1 $\mu$m, as seen from the scanning-electron-micrograph shown in figure~\ref{fig:morphology}.b.
\begin{table}
\begin{center}  
    \begin{tabular}{|p{2.65cm}|c|c|c|c|c|p{1cm}|} \hline
    \multirow{2}{*}{Method} & \multicolumn{5}{|c|}{Position No.} &\multirow{2}{*}{$d \pm \Delta d$} \\
    \cline{2-6}& \multicolumn{1}{|c|} {1} & \multicolumn{1}{|c|} {2} &  \multicolumn{1}{|c|} {3} &  \multicolumn{1}{|c|} {4} &  \multicolumn{1}{|c|} {5}& \\\hline
    
    {Measured using the profilometer ($\mu$m)} & \multicolumn{1}{|c|} {5.3 $\pm$ 0.6} & \multicolumn{1}{|c|} {5.7 $\pm$ 0.9 } & \multicolumn{1}{|c|} {6.1 $\pm$ 1.0} & \multicolumn{1}{|c|} {5.7 $\pm$ 0.9} & \multicolumn{1}{|c|} {6.3 $\pm$ 0.9} & \multicolumn{1}{|c|} {5.8 $\pm$ 0.9} \\ \hline  
    
    {Determined from transmittance ($\mu$m)}& \multicolumn{1}{|c|} {5.8 $\pm$ 0.6} & \multicolumn{1}{|c|} {6.3 $\pm$ 0.7 } & \multicolumn{1}{|c|} {6.3 $\pm$ 0.7} & \multicolumn{1}{|c|} {5.8 $\pm$ 0.6} & \multicolumn{1}{|c|} {5.8 $\pm$ 0.7} & \multicolumn{1}{|c|} {6.0 $\pm$ 0.6}    \\ \hline
    \end{tabular}
\end{center}
\caption{\textbf{Thickness of the PMMA film obtained from two methods:} i) using the profilometer, ii) fitting the FTIR transmittance spectra to the published data \cite{Zhang2020b}. Both measurements, profilometer, and FTIR spectroscopy, are performed at five different locations on the sample surface.
}\label{tab:thickness}
\end{table}

\subsection{Spectrophotometric Measurements}
\label{scpectroscopy}
\begin{figure}[htb]
    \centering\includegraphics[width=0.8\textwidth]{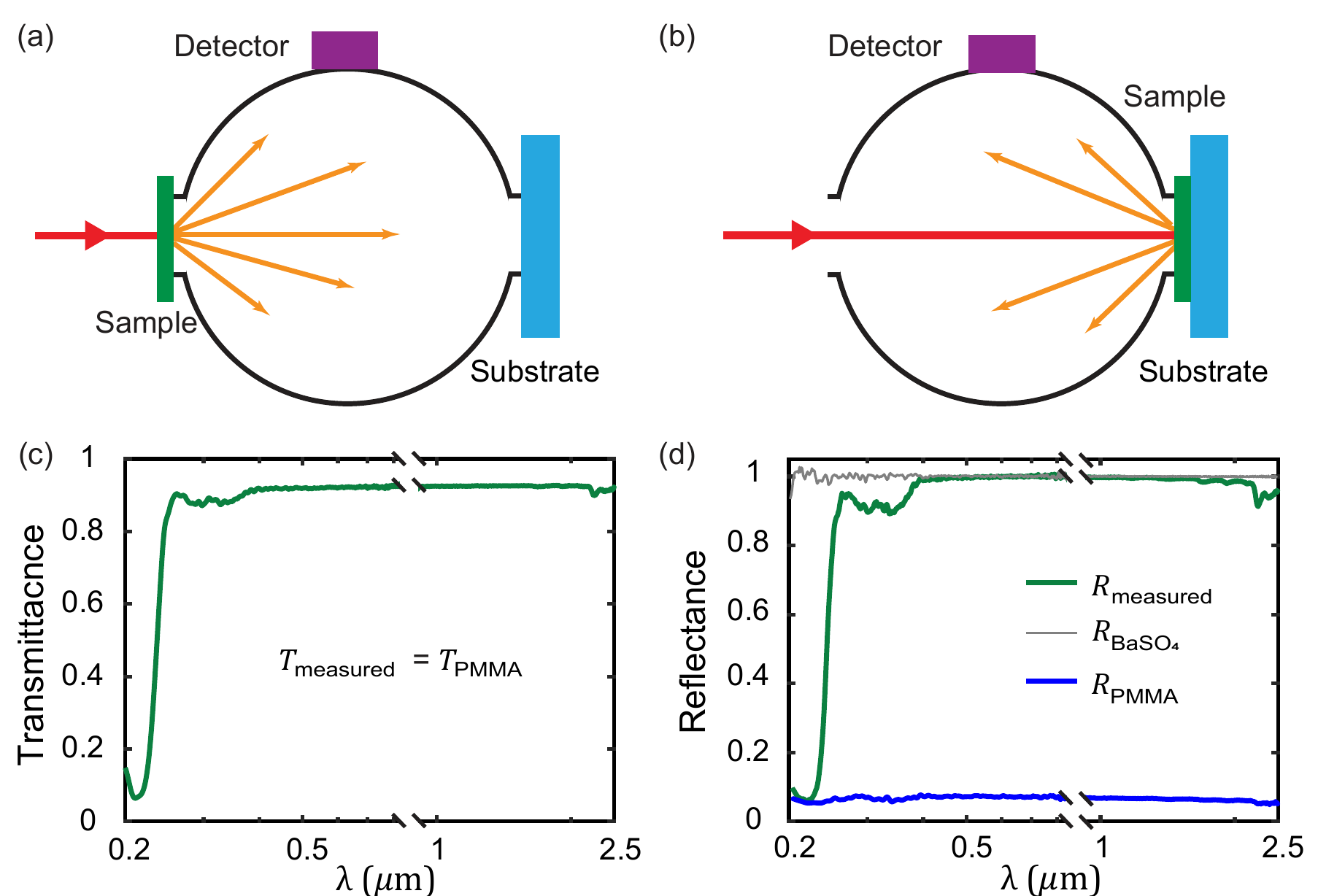}
    \caption{\textbf{(Color online) Measurement diagram and experimental results:} Optical properties of thin PMMA films are measured in a Hitachi UH-4150 spectrophotometer from 0.2 to 2.5 $\mu$m with 0.001 $\mu$m steps with the help of an integrating sphere. (a) The transmittance measurement is performed on the standalone PMMA film. (b) The reflectance is measured with a PMMA film attached to a BaSO$_4$ substrate. (c) The measured result $T_{\rm{measured}}$ in the green (dark-grey) line is the transmittance of the PMMA standalone film. (d) The measured reflectance in the green (dark-grey) line is that of the PMMA film on top of a BaSO$_4$ substrate. The measured reflectance of BaSO$_4$ substrate alone in the light-grey line fluctuates, especially in short-wavelength regions. The reflectance of PMMA film in the blue (black) line is extracted from the measured data by using Eq.~\eqref{eq:Rtotal} in the main text. }\label{fig:meas}
\end{figure}

Optical properties of the PMMA thin film are measured using a standard spectrophotometer in the wavelength from 0.2 to 2.5 $\mu$m with 0.001 $\mu$m step. Figure \ref{fig:meas} represents the measurement setup to obtain the transmittance (a) and reflectance (b) using the Hitachi UH-4150 system equipped with a 60 mm standard integrating sphere. This instrument is equipped with a deuterium lamp in the UV region and a tungsten-halogen lamp in the Vis-NIR region. Since the integrating sphere is coated with a highly-reflective diffuse material on the inner surface, the light reflects numerous times at the inner wall of the integrating sphere before reaching the detector.  
Between 0.2 and 2.5 $\mu$m, the detector switches its wavelength around 0.85 $\mu$m to maintain the sensitivity at the corresponding wavelength. The incidence angles for the regular transmittance and specular reflectance measurements are $0^{\circ}$ and $8^{\circ}$, respectively. {The standard measurement of reflectance at small angles close to the normal incidence angle typically yields only a small deviation, which is less than $0.05\%$ at weak absorption limit and less than $0.1\%$ at strong absorption limit for the specified case in this paper. This is negligible in comparison with the other error discussed later in this section. Therefore, we exclude the effect of a small incidence angle in the extraction of $n$ and $k$.} Both optical measurements are performed with high fidelity. There are unavoidable jumps in the data in this regime \cite{Zhang20}. Therefore we skip the data from 0.81 to 0.91 $\mu$m as shown in the results in figures~\ref{fig:meas}.c and d.

As shown in the diagram in figure~\ref{fig:meas}a, the transmittance is measured on standalone samples. This transmittance is the measured result of the PMMA film as shown in green (dark-grey) in figure~\ref{fig:meas}c. On the other hand, the reflectance is measured following the diagram in figure~\ref{fig:meas}.b. Here, the PMMA film is clamped to a BaSO$_4$ substrate and $R_{\rm{measured}}\neq R_{\rm{PMMA}}$. In figure \ref{fig:meas}.d, the green (dark-grey) line for $R_{\rm{measured}}$ is the total reflectance of both PMMA film and BaSO$_4$ substrate, which depend on each other by\cite{Howell20}
\begin{equation}\label{eq:Rtotal}
R_{\rm{measured}}=R_{\rm{PMMA}}+\frac{R_{\rm{BaSO}_4}T_{\rm{PMMA}}^2}{1-R_{\rm{BaSO}_4}R_{\rm{PMMA}}}.
\end{equation}
This equation is converted into the quadratic equation of the reflectance $R_{\rm{PMMA}}$, which solution is unique due to $R_{\rm{PMMA}}<R_{\rm{measured}}$. With $T_{\rm{PMMA}}$,  $R_{\rm{measured}}$, and $R_{\rm{BaSO}_4}$ $ \approx 1$ as shown in light-grey line in figure \ref{fig:meas}.d, $R_{\rm{PMMA}}$ is determined as shown in the blue (black) line in figure \ref{fig:meas}.d. The reflectance of the standard substrate is expected to be $100\%$\cite{Mikhailov19}. However, there is some deviation,  {which is as large as $6.6\%$}, and we use this deviation to represent the error coming from the measurements. The notations of $R_{\rm{PMMA}}$ and $T_{\rm{PMMA}}$ are only used in this subsection to differentiate them from the measured data. Elsewhere, in the earlier subsection of the reflection-transmission method and the following sections, reflectance and transmittance of PMMA films are denoted as $R$ and $T$, respectively. 

\section{Results and discussion}\label{results}
As seen from the data in figure \ref{fig:meas}. c, the transmittance of the PMMA film exhibits significant changes in different spectral regions due to its chemical compositions. In the ultraviolet regime, the chromophore group $(C=O)$ in the polymer chain gives rise to two electronic excitations \cite{Wu11, Kurapati19}. One excitation arises from the $\pi\Rightarrow\pi^*$ bond, corresponding to the strong absorption in the $0.2-0.27$ $\mu$m range. Hence, the transmittance is as low as $10\%$. The other excitation originates from the $n\Rightarrow\pi^*$ bond. With a weaker absorption, the transmittance is around $90\%$ with a small shoulder in the $0.27-0.38$ $\mu$m range. There is no indication of absorption in the visible region and the large part of the near-infrared region between $0.38$ $\mu$m and $2.0$ $\mu$m. The transmittance is almost flat and approximates 92\%.
In the small part of the near-infrared region from $2.0$ $\mu$m to $2.5$ $\mu$m ranges, the alkyl, -CH$_3$ and -CH$_2$, stretching vibrations give rise to weak absorptions \cite{Forrester13}. The transmittance of PMMA slightly decreases. On the other hand, the reflectance of the standalone PMMA film, figure \ref{fig:meas}.d, varies slightly in $6\%$ - $8\%$ in the whole spectrum since it depends on the reflection at the film surfaces more than the absorption inside the film.

\begin{figure}[htb] 
      \centering
      \includegraphics[width=0.8\textwidth]{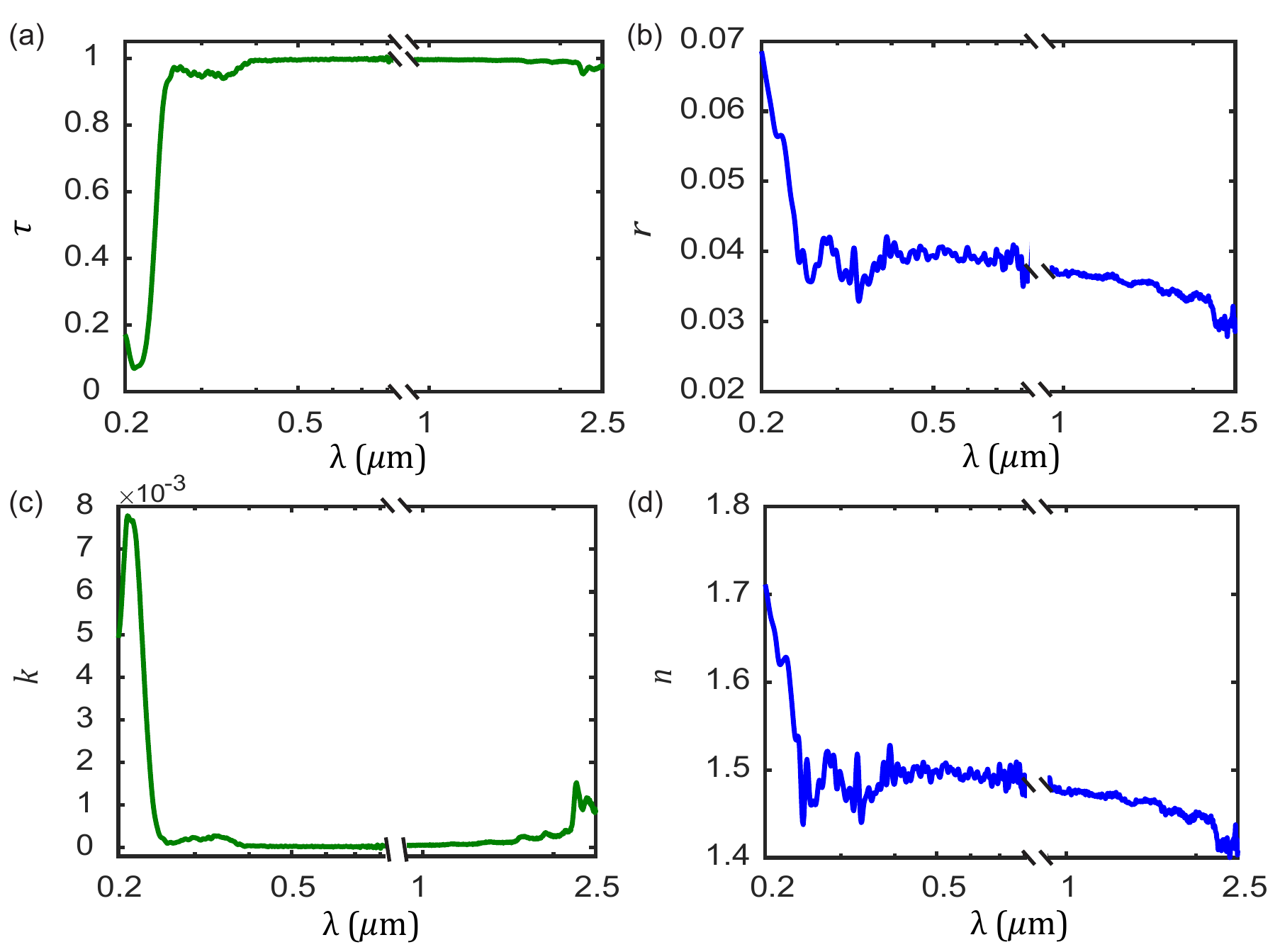}
      \caption{\textbf{(Color online) Optical properties of PMMA film:} (a) the {transmissivity $\tau$} is calculated from Eq.~\eqref{eq:t0}. (b) the reflection coefficient $r$ is calculated from the transmittance $T$, reflectance $R$, and {transmissivity $\tau$} using Eq.~\eqref{eq:r0}. (c) The extinction coefficient $k$ is calculated from Eq.~\eqref{eq:k0}. (d) The refractive index $n$ is calculated using an inverse function from the reflection coefficient $r$ and the extinction coefficient $k$.  
      \label{fig:comp}}
\end{figure}

Using the transmittance and reflectance of PMMA films determined above, we calculate the {transmissivity $\tau$} and reflection coefficient $r$ and show them in figure \ref{fig:comp}.a and figure \ref{fig:comp}.b, respectively. 
In the strong absorption regime from $0.2$ - $0.27$ $\mu$m, a part of the incoming light reflects at the first surface while the transmitted part is absorbed nearly $100\%$ by the film. There is almost no reflection on the second surface. Therefore, $r\approx R$ and $\tau \approx T$. In the weak absorption regime from $0.27$ - $2.5$ $\mu$m, the film has almost no power loss. The {transmissivity $\tau$} is as large as $99\%$. As the reflectance depends on the reflections at both film surfaces, its value at one interface is $r\approx R/2$.

The complex refractive index of PMMA for the full range from $0.2$ to $2.5$ $\mu$m is presented in the bottom row in figure \ref{fig:comp}. The extinction coefficient $k$ and the refractive index $n$ are shown in figure \ref{fig:comp}.c and figure \ref{fig:comp}.d, respectively. From the extinction coefficient $k$, one can see the effect of chemical compositions of materials on optical properties. In the strong-absorption regime from $0.2$ $\mu$m to $0.27$ $\mu$m, the profound peak of $k$ appears to correspond to the excitation from $\pi\Rightarrow\pi^*$ bond. 
In the weak absorption regimes, $k$ shows peaks at $\lambda\approx 0.27$ - $0.38$ $\mu$m corresponding to the excitation from $n\Rightarrow\pi^*$ bond, and at $\lambda\approx 2.0$ - $2.5$ $\mu$m corresponding to alkyl vibration stretchings. These peaks are more significant than the corresponding features in the transmittance in figure \ref{fig:meas}.c. 
Outside the excitation and vibration energy scales of PMMA, $\lambda\approx 0.38$ - $2.0$ $\mu$m, $k$ is close to zero. 
Since the refractive index $n$ is related to the extinction coefficient via the Kramers-Kronig transforms, the changes in $k$ also correspond to the trend in $n$. Therefore, in the ultraviolet regime, the refractive index of PMMA shows a trend of significant change and reaches the value of $1.7$ at $\lambda = 0.2$ $\mu$m. 
In the visible and infrared regions, the refractive index $n$ changes insignificantly and averages $1.47$. In the wavelength range between $2.0$ $\mu$m and $2.5$ $\mu$m, it slightly declines to $1.41$. 

\begin{figure}[ht]
      \centering
      \includegraphics[width=0.8\textwidth]{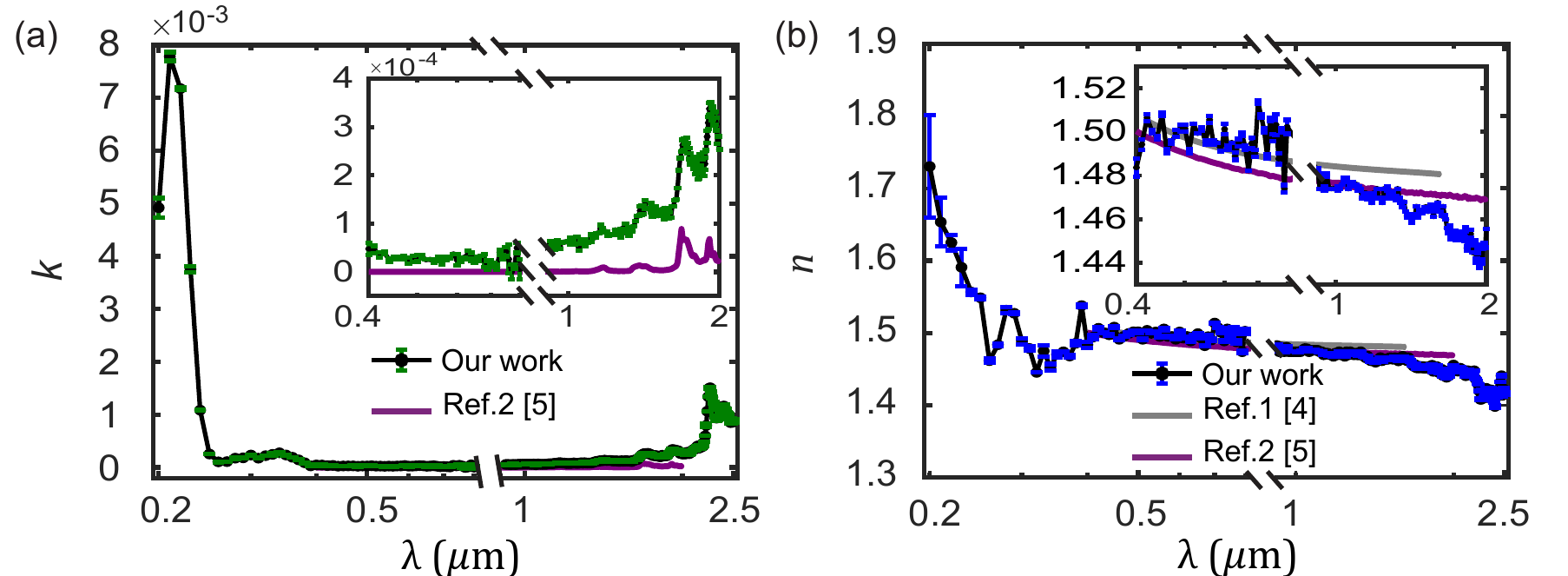}
      \caption{ \textbf{(Color online) The uncertainty of the complex refractive indices:}  (a) extinction coefficient $k$, (b) refractive indices $n$. The values identified are compared to several values of the previous reports as Ref.~\cite{Beadie15}, Ref.~\cite{Zhang20} in the range wavelength from 0.38 $\mu$m to 2 $\mu$m. 
      \label{fig:error}}
\end{figure}

In determining $k$ and $n$ of PMMA, two primary sources affect the accuracy of our final results: instrumental errors and sample uncertainties. In figure~\ref{fig:error}, the uncertainties of complex refractive indices using the error analysis method \cite{Hughes10} are presented as the error bars of the main results. Looking closely into the data set, we found the instrument error contributes exceedingly to the uncertainties of the refractive index $n$ and the extinction coefficient $k$ \cite{pmma}.  
The uncertainty of $k$ shows a strong dependence on the absorption property of the material. At the highest absorption at $\lambda=0.22$ $\mu$m, the relative error of $k$ is less than $1\%$. While, in the lowest absorption-range $0.38<\lambda<0.82$ $\mu$m, where $k$ average value is around $2.5\times 10^{-5}$, the uncertainty of $k$ is even as large as its magnitude. This is well known since the reflection-transmission method is unsuitable for materials with low absorption that $A=1-R-T$ is much lower than the error of the measurements \cite{Li17}. On the contrary, the uncertainty of $n$ hardly depends on the absorption properties as $\Delta n$ varies slightly and is lesser than $4\%$.

In the insets of figure~\ref{fig:error}, the extinction coefficient and the refractive index are compared to those found in the literature \cite{Beadie15,Zhang20}, which are only limited in the Vis-NIR region.  
In the inset of figure \ref{fig:error}.a, the extinction coefficient shows a similar trend to other reported values \cite{Zhang20}. However, the magnitude of the value is somewhat four times larger. Besides the weakness of the reflectance-transmittance method in the strong transmission range mentioned above, it may be due to the $1/f$-- the noise of $PbS$ detector in the infrared spectrum that lifts the result from the correct one \cite{Miroshnikova2010}.
In the inset of figure \ref{fig:error}.b, the refractive index is comparable to those published previously as Ref.~\cite{Beadie15} and Ref.~\cite{Zhang20}. 
The difference between the results of the current study and the previous works, Ref.~\cite{Zhang20, Beadie15}, are relatively small over the full range of 0.4-2.0 $\mu$m. 

As mentioned in subsection~\ref{theory}, the transmittance sensitivity gives a constraint on measuring the extinction coefficient.
In the ultraviolet regime, the extinction coefficient can be obtained with high accuracy up to some constraints depending on the transmission measurement, as shown in Eqs.~\eqref{eq:k0A} and ~\eqref{eq:k0A_2}. 
{From these equations, we can define the maximum value of $k$  that the reflection-transmission method can extract.}
It is known that polymers absorb ultraviolet radiation \cite{Rashidian14, Tuhin20}. As a result, the values of the extinction coefficient cannot be determined with good accuracy in the ultraviolet region because the measured transmission is unreliable when the sample films are too thick \cite{Brissinger19, Tuhin20}. 
This is opposite to the visible and infrared measurements in which the thicker sample gives the more reliable extinction coefficient \cite{Brissinger19, Kasarova07, Sultanova09, Zhang20}. {In this work, we focus more on the ultraviolet range, then the sample is fabricated to be relatively thin, 5.8 $\mu m$.}
With this thickness, our PMMA thin film partly transmits the radiation of wavelength less than 0.28 $\mu$m, as shown in figure \ref{fig:comp}.a. Apparently, the measured transmittance is well above the instrument noise. The estimation of the extinction coefficient yields reliable data in this regime.

\section{Conclusion}
In conclusion, the complex refractive index of Poly(methyl methacrylate) film is established in the UV-Vis-NIR regime from 0.2 $\mu$m to 2.5 $\mu$m. The reflection-transmission method is used to analyze measurement data on reflectance and transmittance and to calculate the complex refractive index. The standalone PMMA films are fabricated using spin-coating at a thickness of $5.8\pm 0.9\mu$m, as measured by a profilometer. The transmittance and reflectance of the film are measured in a standard spectrophotometer using an integrating sphere. The extracted refractive index $n$ in the Vis-NIR regime from 0.38 to 2.5 $\mu$m matches well to previous reports \cite{Rashidian14,Beadie15,Tuhin20,Zhang20}. With a thin enough film that ensures precise transmittance measurements, the extinction coefficient $k$ is determined in the UV regime from 0.2 to 0.38 $\mu$m. Our results provide useful insight into the optical properties of PMMA thin films, especially in the short wavelength regime.

\section*{Acknowledgments.}
Pham Thi Hong was supported by VinBigdata, Vingroup, and supported by the Master, Ph.D. Scholarship Programme/PostDoctoral Scholarship Programme of Vingroup Innovation Foundation (VINIF), Vingroup Big Data Institute (VinBigdata), code VINIF.2021.TS.019.  Samples are fabricated in the cleanroom of the Nano and Energy Center. Optical data are measured at VNU University of Science. Sample thicknesses are measured at Jeonbuk National University. {The calculations in this work are done at the Phenikaa University's HPC Systems.}

\section*{Disclosures.} The author declare no conflicts of interest.

{\section*{Data availability.} Data underlying the results presented in this paper are available in Ref.~\cite{pmma}.}

\bibstyle{unsrt}
\bibliography{reference}

\end{document}